\documentclass[
journal,
twoside,
web,
final,
]{ieeecolor}

\usepackage{algorithmic}
\usepackage{amsmath,amssymb,amsfonts}
\usepackage{bm}
\usepackage[american]{babel}
\usepackage{cite}
\usepackage{dcolumn}
\usepackage{generic}
\usepackage{graphicx}
\usepackage[colorlinks=True,linkcolor=red,citecolor=red,urlcolor=blue]{hyperref}
\usepackage{hhline}
\usepackage[mathlines, pagewise]{lineno}
\usepackage{textcomp}
\usepackage{threeparttable}
\usepackage{setspace}
\usepackage{multirow}
\usepackage{makecell}
\usepackage[version=3]{mhchem}
\usepackage{amsmath}
\usepackage{gensymb}
\usepackage{url}
\usepackage{soul}
\usepackage{textgreek}

\definecolor{red}{rgb}{0.6,0.1,0.1}
\definecolor{blue}{rgb}{0.1,0.1,0.6}
\definecolor{green}{rgb}{0.1,0.6,0.1}
\definecolor{white}{rgb}{1,1,1}
\definecolor{black}{rgb}{0,0,0}

\begin{document}

\title{RESURF \texorpdfstring{\ce{Ga2O3}}{}-on-\texorpdfstring{\ce{SiC}}{} Field Effect Transistors for Enhanced Breakdown Voltage}


\author{Junting Chen, \IEEEmembership{Graduate Student Member, IEEE},
Junlei Zhao,  
Jin Wei, 
and Mengyuan Hua, \IEEEmembership{Member, IEEE}
\thanks{Manuscript received xx Month 2023; revised xx Month 2023; accepted xx Month 2023. Date of publication xx Month 2023; date of current version xx Month 2023. The review of this article was arranged by Editor Xxxx. Yyyyy. \textit{(Corresponding author: Mengyuan Hua.)}}
\thanks{Junting Chen is with the Department of Electrical and Electronic Engineering, Southern University of Science and Technology, Shenzhen 518055, China, and also with the Department of Electronic and Computer Engineering, The Hong Kong University of Science and Technology, Hong Kong, China.}
\thanks{Junlei Zhao, and Mengyuan Hua is with the Department of Electrical and Electronic Engineering, Southern University of Science and Technology, Shenzhen 518055, China (e-mail: huamy@sustech.edu.cn).}
\thanks{Jin Wei is with the Department of Electrical and Electronic Engineering, Peking University, Beijing 100871, China.}
}

\maketitle

\begin{abstract}


Heterosubstrates have been extensively studied as a method to improve the heat dissipation of \ce{Ga2O3} devices. 
In this simulation work, we propose a novel role for $p$-type available heterosubstrates, as a component of a reduced surface field (RESURF) structure in \ce{Ga2O3} lateral field-effect transistors (FETs). 
The RESURF structure can eliminate the electric field crowding and contribute to higher breakdown voltage. 
Using \ce{SiC} as an example, the designing strategy for doping concentration and dimensions of the \textit{p}-type region is systematically studied using TCAD modeling. 
To mimic realistic devices, the impacts of interface charge and binding interlayer at the \ce{Ga2O3}/\ce{SiC} interface are also explored. 
Additionally, the feasibility of the RESURF structure for high-frequency switching operation is supported by the short time constant ($\sim$0.5~ns) of charging/discharging the \textit{p}-SiC depletion region. 
This study demonstrates the great potential of utilizing the electrical properties of heat-dissipating heterosubstrates to achieve a uniform electric field distribution in \ce{Ga2O3} FETs.

\end{abstract}


\section{Introduction} \label{sec:introduction}

\IEEEPARstart{G}{allium} oxide (\ce{Ga2O3}) has attracted intensive research interests in deep-ultraviolet photodetectors~\cite{ Pearton2018a}, power electronics~\cite{green2022gallium, Zhang2022Ultra}, and radio-frequency applications~\cite{singh2018pulsed} in recent years. 
With the merit of intrinsic material properties of \ce{Ga2O3}, such as ultrawide bandgap ($E_\mathrm{g} \simeq$ 4.8~eV), high critical electric field ($E_\mathrm{crit.} \simeq$ 8~MV$\cdot$cm), and decent electron mobility ($\mu_\mathrm{n} \simeq$ 200~cm$^{2}$$\cdot$V$^{-1}$$\cdot$s$^{-1}$), the \ce{Ga2O3}-based power electronics have much higher Baliga’s figure of merit (BFOM) compared to Si and other wide-bandgap semiconductors~\cite{green2022gallium}. 
However, the intrinsically low thermal conductivity of \ce{Ga2O3} ($\kappa_\mathrm{T} \simeq$ 0.1-0.3~W$\cdot$m$^{-1}$$\cdot$K$^{-1}$~\cite{santia2015lattice}) becomes a major obstacle in the design of power electronics, because the resultant 
high thermal resistance and self-heating effects seriously hamper the efficiency, reliability, and scalability of the \ce{Ga2O3} power devices~\cite{zhang2021how}, and hence their competitiveness in power applications.

As a remedy for this thermal management challenge, heat dissipation of \ce{Ga2O3}-based devices can be improved by utilizing heterogeneous semiconductor substrates with higher thermal conductivity, such as sapphire~\cite{hu2018lateral, polyakov2019deep}, diamond~\cite{wang2022high, cheng2020integration, cheng2019thermal}, \ce{GaN}, and \ce{SiC}.
Among these heterosubstrates, \ce{SiC} is an attractive choice owing to its high thermal conductivity and low cost. 
The thermal conductivity of \ce{SiC} is 490~W$\cdot$m$^{-1}$$\cdot$K$^{-1}$, better than most typical semiconductor substrates (e.g., 150~W$\cdot$m$^{-1}$$\cdot$K$^{-1}$ for Si, 46~W$\cdot$m$^{-1}$$\cdot$K$^{-1}$ for sapphire, and 130~W$\cdot$m$^{-1}$$\cdot$K$^{-1}$ for \ce{GaN}~\cite{Song2021Ga2O3}). 
Compared with diamond, large-diameter \ce{SiC} wafers can be produced at a much lower cost which can be further reduced with polycrystalline samples~\cite{lin2019single}.
Heterogeneous binding of \ce{Ga2O3} on \ce{SiC} substrate has been demonstrated through surface activation bonding~\cite{lin2019single}, fusion bonding~\cite{Song2021Ga2O3}, and direct growth~\cite{Hrubisak2023Heteroepitaxial, Nepal2020Heteroepitaxial}. 
Moreover, wafer-scale integration of hundreds-nm \ce{Ga2O3} thin film on \ce{SiC} can be achieved with the ion-cutting technique~\cite{xu2019first}. 

As a result of the effectively promoted heat dissipation, \ce{Ga2O3} field-effect transistors (FETs) based on such heterosubstrates deliver improved thermal stability and suppressed self-heating effects~\cite{Mahajan2019Electrothermal, Russell2017Heteroepitaxial}.
Currently, overcoming the interfacial thermal impediments of the heterogeneous integration is still an active area of research. 
However, the investigations are mainly focused on the thermal properties and heat dissipation, whereas the electrical properties of the heterosubstrates and their potential benefits for the power electronics have not been extensively explored.

\begin{figure*}[ht!] 
 \includegraphics[width=18.1cm]{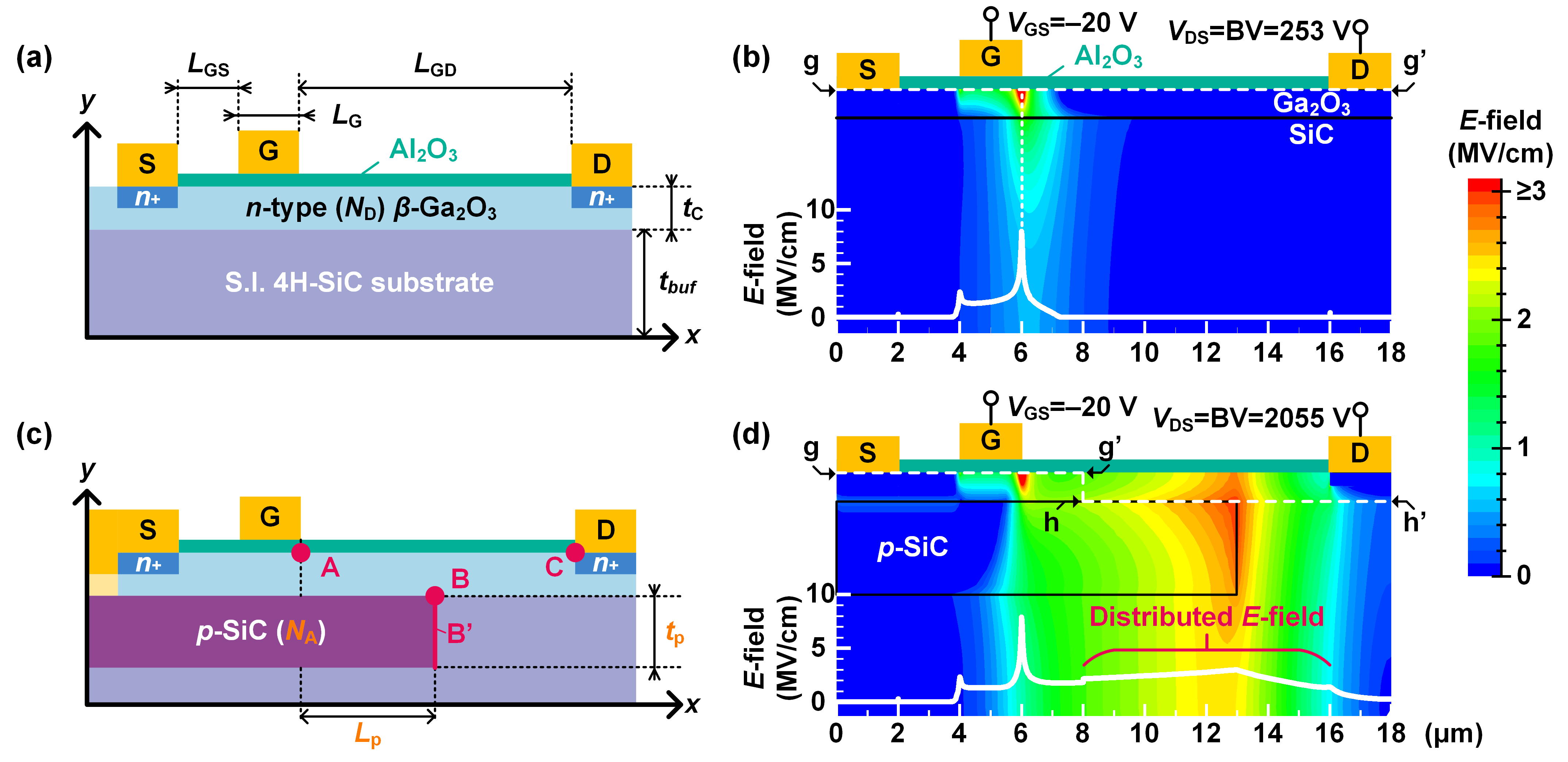}
 \caption{Schematic structures and \textit{E}-field strengths at breakdown voltages of (a,b) conventional \ce{Ga2O3}/\ce{SiC} FET; and (c,d) FET with the \textit{n}-\ce{Ga2O3}/\textit{p}-\ce{SiC} RESURF structures. 
 The device with RESURF structures withstands much higher \textit{V}\textsubscript{DS}.
 In (b), the \textit{E}-field along cutline g-g' is shown in the inset. In (d), the \textit{E}-field along cutline g-g' and h-h' is shown in the inset.
 The red dots (A, B, and C) and red line (B') in (c) are potential breakdown positions, and are critical in the following discussion.
 The values of key simulation parameters in (a): \textit{L}\textsubscript{GS}, \textit{L}\textsubscript{G}, \textit{L}\textsubscript{GD}, \textit{t}\textsubscript{c}, \textit{t}\textsubscript{buf}, \textit{N}\textsubscript{D} are 2~\textmu m, 2~\textmu m, 10~\textmu m, 0.2~\textmu m, 1.5~\textmu m, and 3$\times$10\textsuperscript{17}~cm\textsuperscript{-3}, respectively.
 One of the optimal values of \textit{N}\textsubscript{A}, \textit{L}\textsubscript{p}, \textit{t}\textsubscript{p} that used in (d) is 0.6$\times$10\textsuperscript{17}~cm\textsuperscript{-3}, 7~\textmu m, 0.65~\textmu m, respectively.
 The breakdown is defined as the \textit{E}-field in the \ce{Ga2O3} reaches 8~MV$\cdot$cm\textsuperscript{-1}, or the \textit{E}-field in the \ce{SiC} reaches 3~MV$\cdot$cm\textsuperscript{-1}.
 }
 \label{fig:fig1}
\end{figure*}

In this work, we demonstrate the utilization of $p$-type doping of \ce{SiC} heterosubstrate as a component of a reduced surface field (RESURF) structure to suppress electric field ($E$-field) crowding in lateral \ce{Ga2O3} FETs. 
The $E$-field crowding is one of the major issues that limit the breakdown voltages (BVs) of lateral high-voltage devices. 
The RESURF technology is firstly developed to address this issue in silicon devices~\cite{Appels1979High}, and is later implemented in \ce{SiC}~\cite{Saks1999high, Chatty2000High}, and \ce{GaN} devices~\cite{Karmalkar2001RESURF}, showing remarkable benefits in terms of more evenly-distributed $E$-field, and thus higher BVs.
As for \ce{Ga2O3}, the lack of $p$-type doping techniques poses a fundamental challenge to realizing the RESURF structures. 
In contrast, $p$-type doping is readily available for \ce{SiC} substrates. The upper limit of the acceptor doping concentration in \ce{SiC} is as high as $10^{20}$ cm$^{-3}$, with the feasibility of selective-area doping. 
Therefore, in addition to improving the thermal dissipation, the \ce{Ga2O3}-on-\ce{SiC} heterosubstrate also offers an excellent platform to construct RESURF structures with $n$-type \ce{Ga2O3} to achieve a more evenly distributed $E$-field, and thus higher BV.
To guide the experimental demonstration of the RESURF structures,
the impact of key parameters, such as acceptor concentration ($N_\mathrm{A}$) and dimensions of the $p$-\ce{SiC} region are studied.
To mimic the realistic devices, the influences of \ce{Al2O3} binding interlayer and charges at the \ce{Ga2O3}/\ce{SiC} interface are investigated. 
In final, the charging/discharging rates of the $p$-\ce{SiC} region during the fast switching transient are also investigated.

\section{Proposal of RESURF Structures and Fabricating Feasibility} \label{sec:2}

As shown in Fig.~\ref{fig:fig1}, the device structure (Fig.~\ref{fig:fig1}~(a)) and the $E$-field distribution (Fig.~\ref{fig:fig1}~(b)) of the conventional heterosubstrate \ce{Ga2O3}-on-\ce{SiC} FET is compared with that of a proposed selective-area $p$-\ce{SiC} RESURF FET (Fig.~\ref{fig:fig1}~(c)~and~(d)).
In the conventional FET, the $E$-field crowds at the drain-side gate corner, and the BV of the device is 253~V. 
By introducing a selective-area $p$-\ce{SiC} with optimal parameters (Fig.~\ref{fig:fig1}~(d)), the $E$-field is distributed over a larger area instead of crowding at the gate corner, benefiting to a higher BV of 2055~V.
It is noticed that, the RESURF structures can be combined with other techniques (e.g., field plates~\cite{Sharma2020Field}, and $p$-\ce{NiO}~\cite{Wang2021first} on the top surface) to further increase the BV of devices.

The fabrication of the proposed RESURF \ce{Ga2O3}-on-\ce{SiC} FETs is feasible based on existing techniques. 
The conventional \ce{Ga2O3}-on-\ce{SiC} FETs have been successfully fabricated~\cite{xu2019first, Wang2021Channel, Song2023Ultra}.
Based on the existing devices, the fabrication of the proposed RESURF FETs requires two additional steps: (i) the selective area $p$-doping of the \ce{SiC} substrate, and (ii) the Ohmic contact of the $p$-\ce{SiC}.
The selective-area $p$-doping technique is well-established and commonly used in \ce{SiC} devices~\cite{Godignon2020Silicon}. 
Moreover, the doping of the \ce{SiC} substrate can be done before integrated to the \ce{Ga2O3} layer, so the engineering of $p$-\ce{SiC} region is rather independent from the following fabrication of \ce{Ga2O3} devices.
As for the Ohmic contact, forming Ohmic contact to the $p$-\ce{SiC} needs high temperature (600-1100 \celsius) annealing process~\cite{Huang2020critical}, whereas forming Ohmic contact to the $n$-\ce{Ga2O3} needs low temperature (400-600 \celsius) annealing process~\cite{Lee2021Process}. 
As a result, the Ohmic contact to the $p$-\ce{SiC} should be formed before the formation of the Ohmic contact to the $n$-\ce{Ga2O3}. 
Overall, utilizing the electrical properties of the heterosubstrate by the proposed RESURF structures would not add too much complexity to the device fabrication process.

\begin{figure}[t] 
 \includegraphics[width=8.9cm]{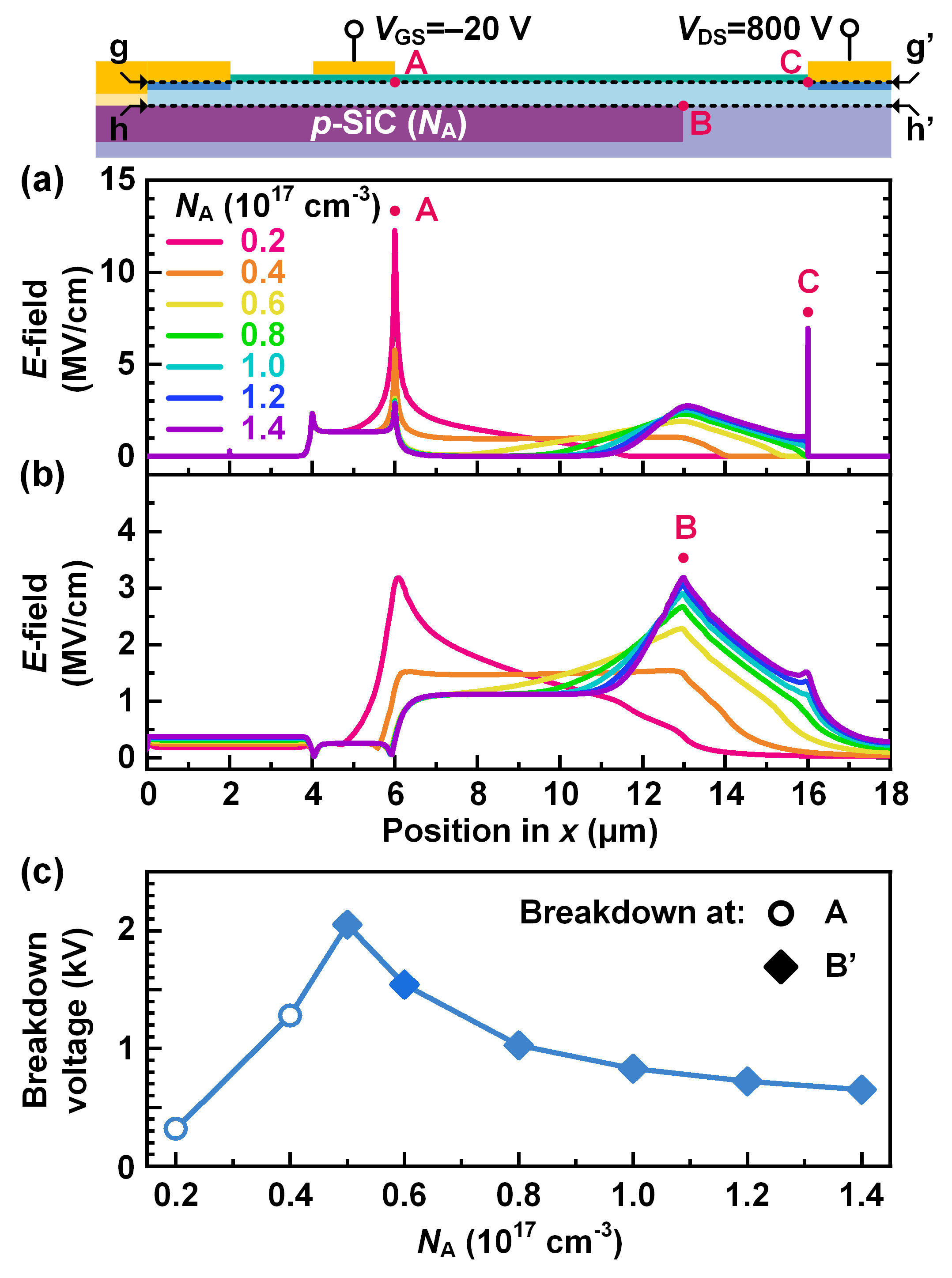}
 \caption{
\textit{E}-field strength along cutline (a) g-g' and (b) h-h' of FET with the \textit{n}-\ce{Ga2O3}/\textit{p}-\ce{SiC} RESURF structure with increasing \textit{N}\textsubscript{A} at \textit{V}\textsubscript{DS} = 800~V. 
Fixed \textit{L}\textsubscript{p} = 7 \textmu m and \textit{t}\textsubscript{p} = 0.8 \textmu m
(c) Overall BVs depending on the \textit{N}\textsubscript{A}.
 }
 \label{fig:fig3}
\end{figure}

\begin{figure}[t] 
 \includegraphics[width=8.9cm]{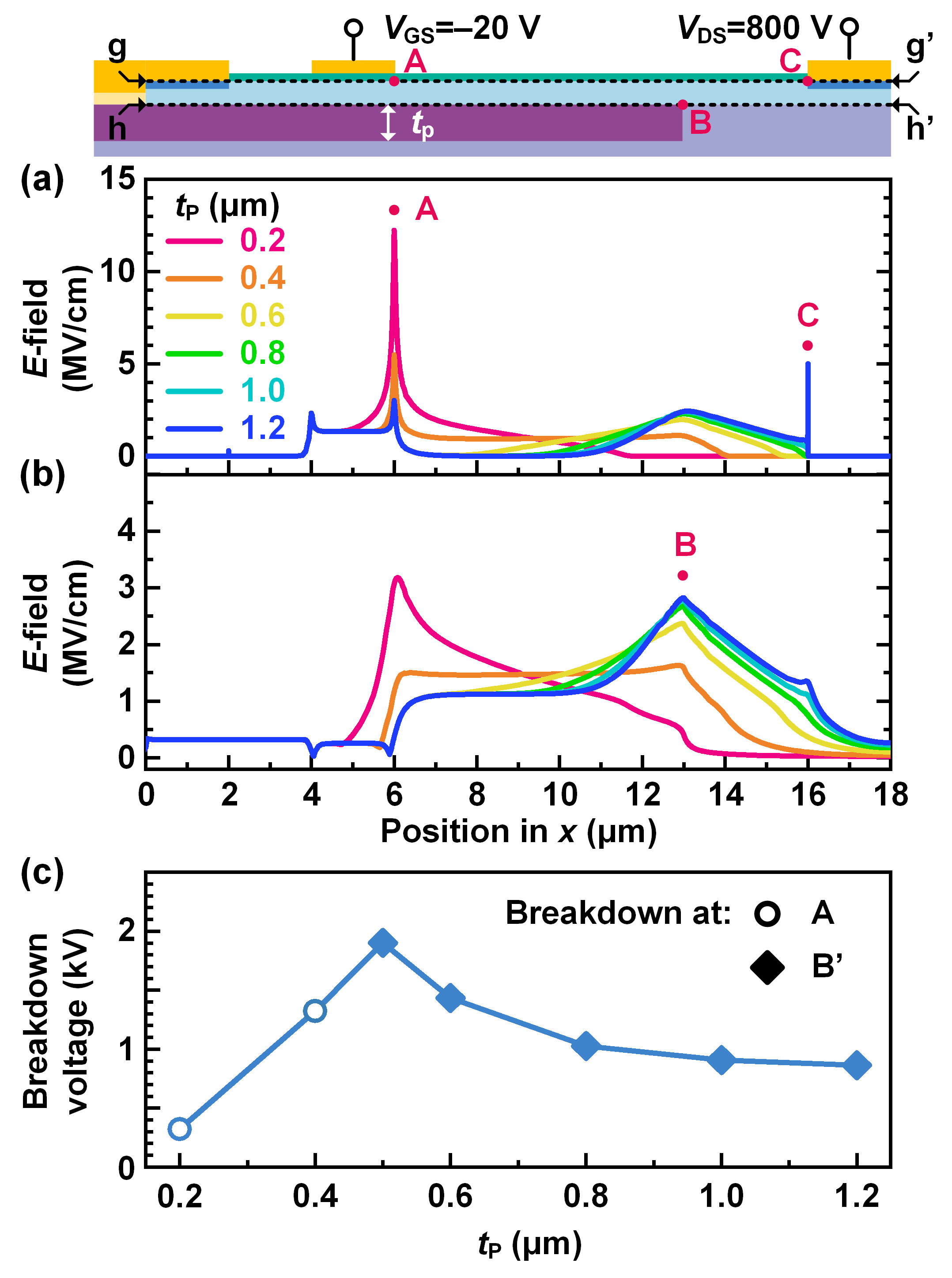}
 \caption{
\textit{E}-field strength along cutline (a) g-g' and (b) h-h' of FET with the \textit{n}-\ce{Ga2O3}/\textit{p}-\ce{SiC} RESURF structure with changing \textit{t}\textsubscript{p} at \textit{V}\textsubscript{DS} = 800~V. 
Fixed \textit{N}\textsubscript{A} = 0.8$\times$10\textsuperscript{17} cm\textsuperscript{-3} and \textit{L}\textsubscript{p} = 7 \textmu m.
(c) Overall breakdown voltages depending on the \textit{t}\textsubscript{p}.
  } 
 \label{fig:fig4}
\end{figure}

\section{\textit{E}-field Regulation in RESURF Structures} \label{sec:3}

In this section, the impacts of the three key parameters, including the acceptor concentration ($N_\mathrm{A}$), the thickness ($t_\mathrm{p}$), and the length ($L_\mathrm{p}$) of the $p$-\ce{SiC} region (as illustrated in Fig.~\ref{fig:fig1}~(c)) are systemically studied.
Meanwhile, the influence of the interface charges and the \ce{Al2O3} interlayer at the \ce{Ga2O3}/\ce{SiC} interface in practical devices are also discussed.
In the following discussions, as labeled in Fig.~\ref{fig:fig1}~(c), three critical potential breakdown positions are: (i) the gate corner at drain side (point~A), (ii) the $p$-\ce{SiC} edge at drain side (line~B'), and (iii) the corner of the drain-side $n^{+}$-\ce{Ga2O3} (point~C).
We note that the $E$-field at point~B is very close or equal (97-100\% at $V_\mathrm{DS}$=800~V) to the peak value of the $E$-field along line~B'.
Therefore, for simplicity, the $E$-field at point~B will be used to represent that along line~B'.

\subsection{Acceptor Concentration of the \textit{p}-SiC} 

Fig.~\ref{fig:fig3}~(a) and (b) show the $E$-field distribution along the cutlines g-g' and h-h', respectively, with the different $N_\mathrm{A}$ in the $p$-\ce{SiC} at a $V_\mathrm{DS}$ of 800~V.
Without the RESURF structures, the $E$-field tends to crowd at point~A, leading to low-voltage device breakdown.
Low-level $p$-doping ($N_\mathrm{A}$ in range of 0.2-0.6$\times10^{17}$~cm$^{-3}$) contributes negative space charge in the $p$-\ce{SiC} to terminate a part of the $E$-field that originally directs to point~A, and results in a fast reduction of the $E$-field at point A from 12.2~MV$\cdot$cm$^{-1}$ to 3.1~MV$\cdot$cm$^{-1}$.
However, as a compensation, the $E$-field at point B increases from 0.4~MV$\cdot$cm$^{-1}$ to 2.3~MV$\cdot$cm$^{-1}$. 
With the $N_\mathrm{A}$ further increasing to 1.4$\times10^{17}$~cm$^{-3}$, more holes in the $p$-\ce{SiC} can further deplete the \ce{Ga2O3} channel.
Consequently, the depletion region in the \ce{Ga2O3} extends toward drain terminal, leading to the $E$-field at point~C increases from 0.4~MV$\cdot$cm$^{-1}$ to 6.9~MV$\cdot$cm$^{-1}$, and that at point~B further increases to 3.2~MV$\cdot$cm$^{-1}$.
Meanwhile, the $E$-field at point~A shows less significant decrease (2.7~MV$\cdot$cm$^{-1}$ at $N_\mathrm{A}$ of 1.4$\times10^{17}$~cm$^{-3}$).

\begin{figure}[t] 
 \includegraphics[width=8.9 cm]{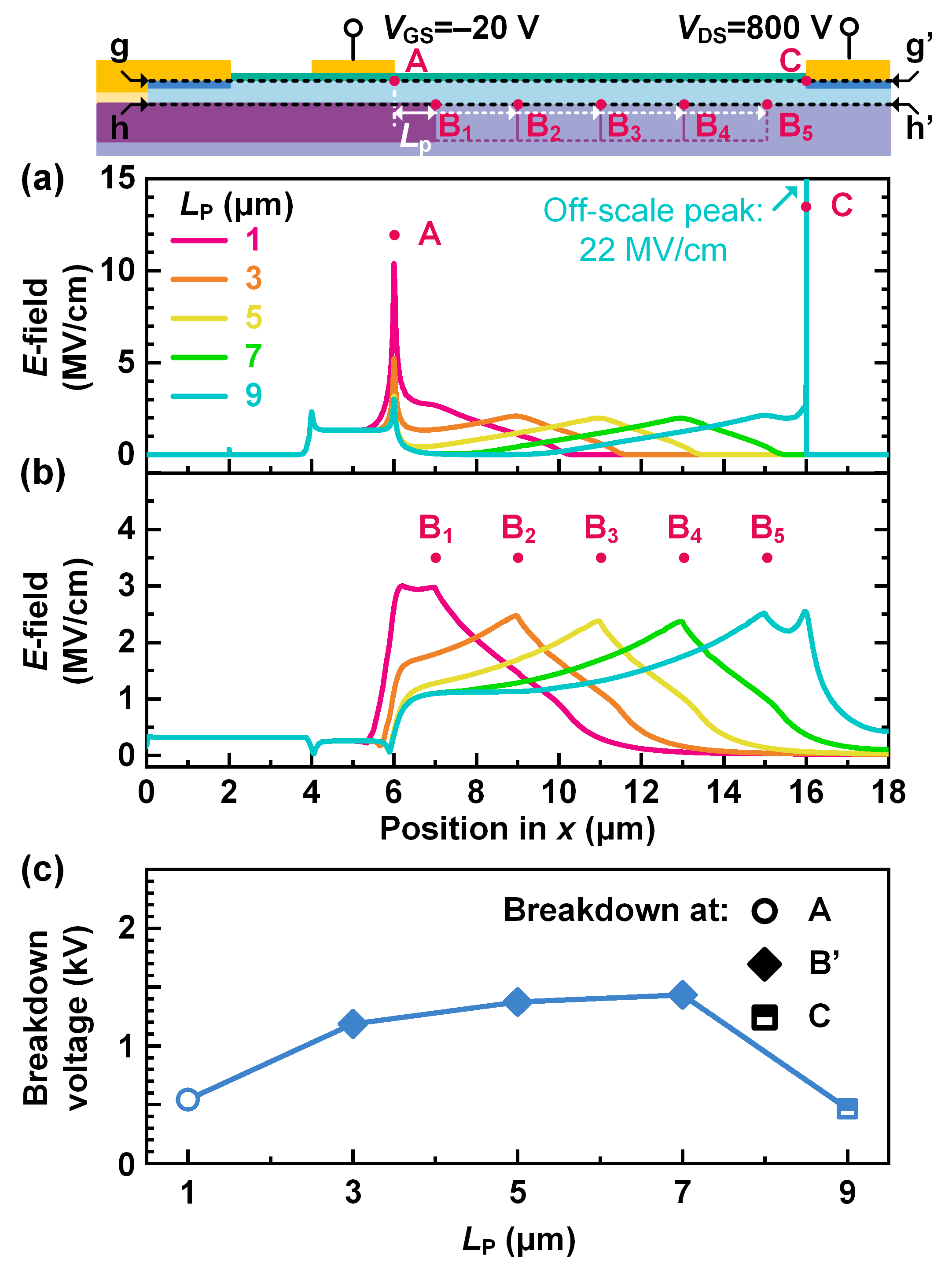}
 \caption{
  \textit{E}-field strength along cutline (a) g-g' and (b) h-h' of FET with the \textit{n}-\ce{Ga2O3}/\textit{p}-\ce{SiC} RESURF structure with changing \textit{L}\textsubscript{p} at \textit{V}\textsubscript{DS} = 800~V. 
  Fixed \textit{N}\textsubscript{A} = 0.8$\times$10\textsuperscript{17} cm\textsuperscript{-3} and \textit{t}\textsubscript{p} = 0.6 \textmu m.
  (c) Overall BVs depending on the \textit{L}\textsubscript{p}.
  }
 \label{fig:fig5}
\end{figure}

\begin{figure}[t] 
 \includegraphics[width=8.9 cm]{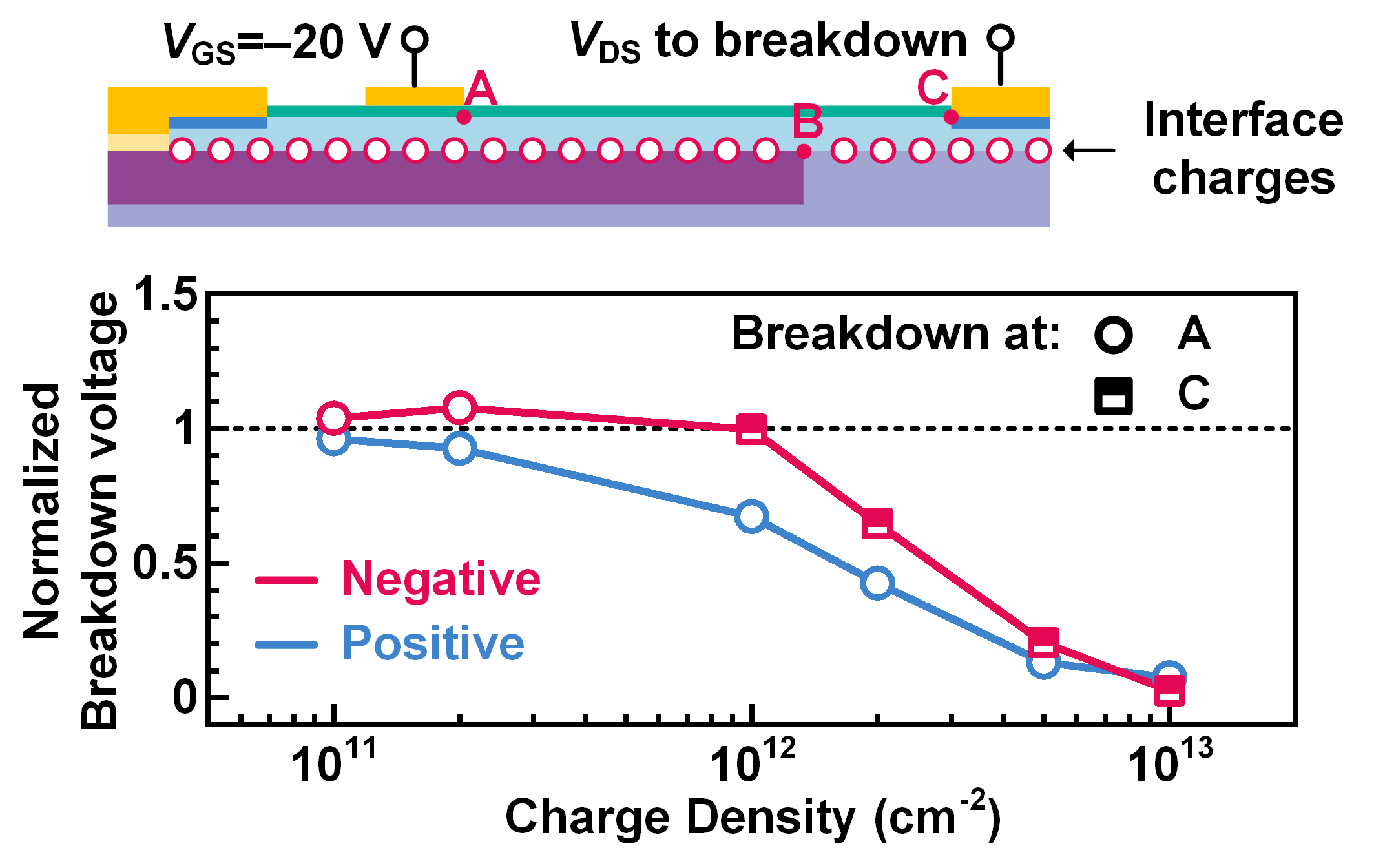}
 \caption{Effects of the different types of charge at the \ce{Ga2O3}/\ce{SiC} interfaces on the BVs.
 Fixed \textit{N}\textsubscript{A} = 0.6$\times$10\textsuperscript{17} cm\textsuperscript{-3}, \textit{t}\textsubscript{p} = 0.6 \textmu m, and \textit{L}\textsubscript{p} = 7 \textmu m.
 }
 \label{fig:fig6}
\end{figure}

Since line~B' ($p$-\ce{SiC}) has a lower critical $E$-field (3~MV$\cdot$cm$^{-1}$) compared to point~A (\ce{Ga2O3}, 8~MV$\cdot$cm$^{-1}$), the breakdown point can easily transfer from point~A to line~B' when the $p$-\ce{SiC} starts to terminate the $E$-field.
Fig.~\ref{fig:fig3}~(c) shows the relationship between the overall BV and $N_\mathrm{A}$. 
When the $N_\mathrm{A}$ increases from 0.2$\times10^{17}$ to 0.5$\times10^{17}$~cm$^{-3}$, the BV increases because the $E$-field at point~A is reduced. 
Starting from the $N_\mathrm{A}$ of 0.5$\times10^{17}$~cm$^{-3}$, the breakdown at line~B' dominates the breakdown of the devices. 
The BV decreases with the further increasing of $N_\mathrm{A}$ due to the increase of the $E$-field along line~B'. 
Overall, the $N_\mathrm{A}$ of 0.5$\times10^{17}$~cm$^{-3}$ is the nearly balanced value for the $E$-field at points~A and B to simultaneously reach their critical $E$-fields.

\subsection{Thickness of the \textit{p}-SiC}

As shown in the Fig.~\ref{fig:fig4}~(a) and (b), the effect of $t_\mathrm{p}$ on the $E$-field distribution is similar to that of $N_\mathrm{A}$.
As expected, the increase of $t_\mathrm{p}$ firstly redirects a portion of the $E$-field from point~A to point~B, indicated as a fast reduction of $E$-field at point~A.
Moreover, the further increase of $t_\mathrm{p}$ extends the depletion region of the $n$-\ce{Ga2O3} channel toward drain terminal, resulting in the larger $E$-field from point~C to point~B. 
Similar to the overall trend in Fig.~\ref{fig:fig3}~(c), the BV increases with $t_\mathrm{p}$ increasing from 0.2~\textmu m to 0.5~\textmu m (Fig.~\ref{fig:fig4}~(c)), owning to the reduction of $E$-field peak at point~A. 
Then, the BV decreases with the $t_\mathrm{p}$ further increasing from 0.5~\textmu m to 1.2~\textmu m owing to the increase of the $E$-field peak at point~B.

Based on the analyses shown in Figs.~\ref{fig:fig3} and \ref{fig:fig4}, we conclude that an optimal overall BV of the device can be reached by carefully engineering the $N_\mathrm{A}$ and $t_\mathrm{p}$ in the $p$-\ce{SiC} region with respect to a balanced $E$-field distribution simultaneously close to the critical points of the $E$-field at point~A in \ce{Ga2O3} and point~B in \ce{SiC}.
It is worth noting that, in both Fig.~\ref{fig:fig3}~(a) and Fig.~\ref{fig:fig4}~(a), $E$-field peaks appear at point~C when the $N_\mathrm{A}$ and $t_\mathrm{p}$ is very large (at $N_\mathrm{A}$ of 1.4$\times10^{17}$~cm$^{-3}$ and $t_\mathrm{p}$ of 1.2~$\mu$m, respectively).
However, in those cases, the line~B' in the \ce{SiC} breakdowns before the $E$-field at point~C reaches the critical value. 
As a result, the $E$-field at point~C would not be a critical concern as long as $N_\mathrm{A}$ and $t_\mathrm{p}$ is not aggressively large.

\subsection{Length of the \textit{p}-SiC}

Fig.~\ref{fig:fig5}~(a) and (b) show the $E$-field distribution with different $L_\mathrm{p}$. 
The $p$-\ce{SiC} region with too short $L_\mathrm{p}$ can not sufficiently block the $E$-field from crowding at point~A.
An $E$-field peak of 8.5~MV$\cdot$cm$^{-1}$ can be seen at the $L_\mathrm{p}$ of 1~\textmu m.
On the other hand, the $p$-\ce{SiC} should not be too close to the drain terminal, otherwise it will cause the $E$-field to crowd at point~C.
A sudden increase of the $E$-field peak at point~C (22~MV$\cdot$cm$^{-1}$) can be found when the $L_\mathrm{p}$ reaches 9~\textmu m (Fig.~\ref{fig:fig5}~(a)).
When the $L_\mathrm{p}$ is in a proper range (3-7~\textmu m in this simulation), the $E$-field at point~A and point~B decreases monotonically with $L_\mathrm{p}$ increase (Fig.~\ref{fig:fig5}~(b)), because a longer $p$-\ce{SiC} benefits to a longer depletion region in the \ce{Ga2O3} channel, and thus a smaller $E$-field at a same $V_\mathrm{DS}$ bias.
To sum up, an $L_\mathrm{p}$ of 7~\textmu m is the optimal $L_\mathrm{p}$ in this simulation, which provides a wide depletion region in \ce{Ga2O3} channel, but does not lead to $E$-field crowing at point~C.

\subsection{Interface Charge}

To mimic the realistic condition of device, the impact of net charges at the \ce{Ga2O3}/\ce{SiC} interface is studied.
In Fig.~\ref{fig:fig6}, the BV with respect to the interface charge density is plotted.
The BV is normalized to that of the device without interface charge. 
A small amount of negative charge (2$\times$10$^{11}$~cm$^{-2}$) can act as field plates, which help blocking the point~A from high $E$-field, thus slightly improve the breakdown voltage.
As for the case where there is a large amount of negative interface charge, it is similar to the case where the $L_\mathrm{p}$ is very long (9~\textmu m in Fig.~\ref{fig:fig5}).
Excessive negative interface charge will cause the $E$-field to crowd at point~C, lowering the breakdown voltage.

Positive interface charge has monotonic impact on the breakdown voltage.
The positive interface charge compensates the negative space charge in the $p$-\ce{SiC} region, and thus eliminates the improvement brought by the RESURF structures, causing the $E$-field to crowd at point~A again.
Moreover, the excessive positive interface charge can emit $E$-field to point~A, further enhancing the $E$-field at point~A.
As a result, the BV decreases monotonically with positive interface charge. 
In practical applications, the interface charge at the \ce{Ga2O3}/\ce{SiC} interface should be carefully controlled.

\subsection{\texorpdfstring{\ce{Al2O3}}{} Interlayer}

\begin{figure}[t] 
 \includegraphics[width=8.9 cm]{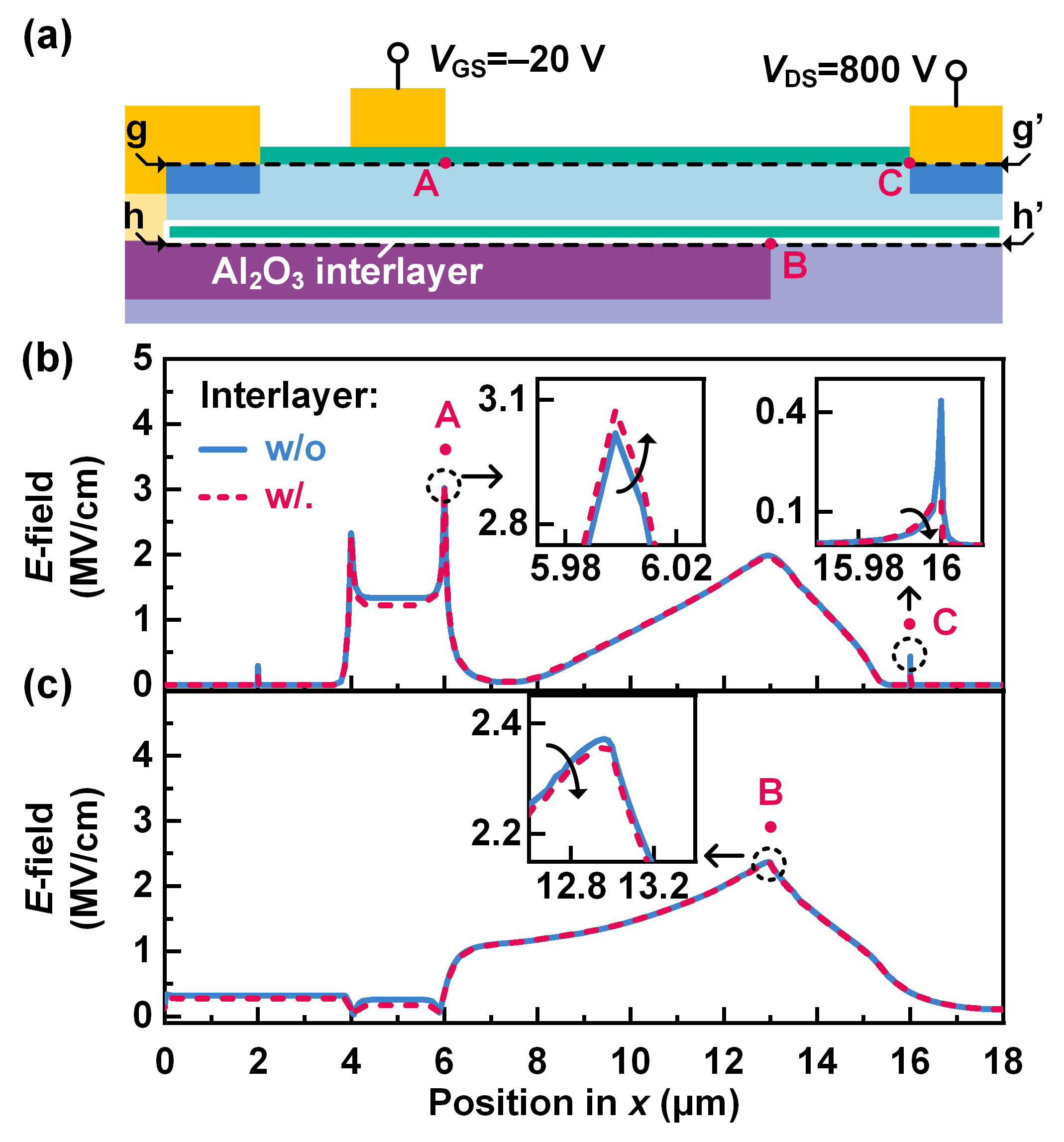}
 \caption{ (a) Schematic illustration of \ce{Al2O3} 
  interlayer. 
  \textit{E}-field strength along the cutlines (b) g-g' and (c) h-h' of the device with and without the 20-nm \ce{Al2O3} interlayer.
  Fixed \textit{N}\textsubscript{A} = 0.8$\times$10\textsuperscript{17} cm\textsuperscript{-3}, \textit{t}\textsubscript{p} = 0.8 \textmu m, and \textit{L}\textsubscript{p} = 7 \textmu m. \textit{V}\textsubscript{DS} = 800 V. 
 }
 \label{fig:fig7}
\end{figure}

An additional \ce{Al2O3} interlayer at the \ce{Ga2O3}/\ce{SiC} interface (Fig.~\ref{fig:fig7}~(a)) is often adopted to improve the interface binding quality and hence the thermal dissipation~\cite{xu2019first, Wang2021Channel}. 
The $E$-field distribution with and without the 20-nm \ce{Al2O3} interlayer is compared in Fig.~\ref{fig:fig7}~(b) (cutline g-g') and Fig.~\ref{fig:fig7}~(c) (cutline h-h'). 
The \ce{Al2O3} interlayer slightly weakens the impact of the RESURF structure, leading to a 1.7\% increase of peak $E$-field at point~A, a 0.4\% decrease of peak $E$-field at point~B, and a 67\% decrease of peak $E$-field at point~C.
As points~A and B are the dominate breakdown points in proper designed devices, and the influence of the \ce{Al2O3} interlayer at points~A and B is quite small, the \ce{Al2O3} interlayer in the practical devices will not limit the adoption of the proposed RESURF structures.

In summary, with the preset dimensions and doping concentration of the $n$-\ce{Ga2O3} channel epilayers, the optimal BV can be obtained when the point~A and the line~B' reach their critical $E$-field simultaneously.
In the example shown in Fig.~\ref{fig:fig1}~(d), by carefully engineering the $N_\mathrm{A}$ and the dimensions of $p$-\ce{SiC}, at $V_\mathrm{DS}$=BV=2055~V, the $E$-field at point~A is 7.91~MV$\cdot$cm$^{-1}$, and that at point~B is 2.99~MV$\cdot$cm$^{-1}$, indicating that point~A and point~B breakdown almost at the same time.
Meanwhile, the interface charges should be minimized in practical fabrication process, and the binding interlayer has less impact on the BV.

\section{Switching Speed of RESURF Devices} 
\label{sec:4}

\begin{figure}[t] 
 \includegraphics[width=8.9 cm]{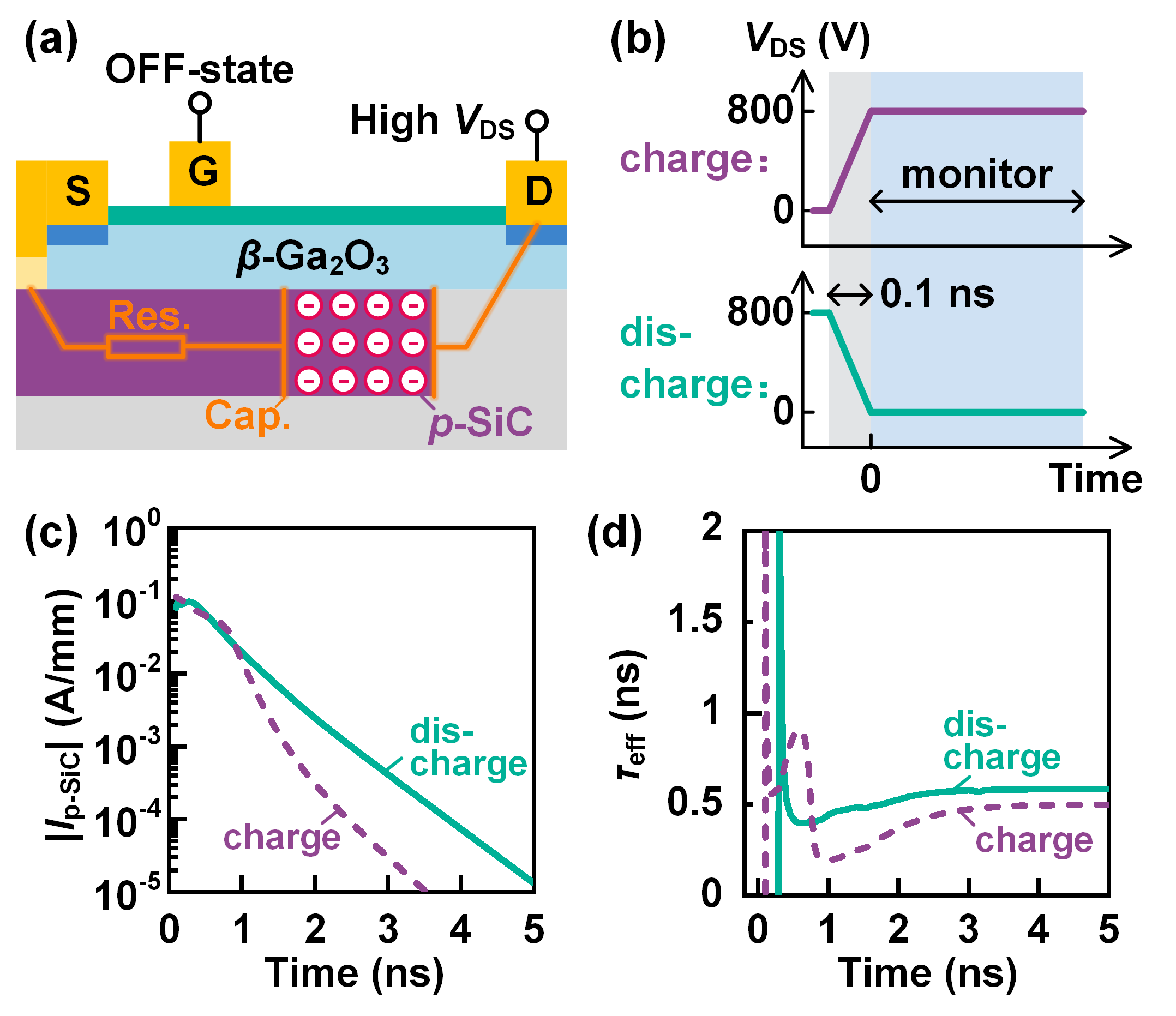}
 \caption{ 
 (a) Schematic illustration of the depletion region in the \textit{p}-\ce{SiC} at high \textit{V}\textsubscript{DS} bias, and its charging/discharging path. "Cap." represents the capacitor formed by the depletion region. "Res." is the equivalent resistance when charging/discharging the depletion region.
 (b) Illustration of the \textit{V}\textsubscript{DS} waveform applied in the simulation. 
 (c) Absolute current passing through the \textit{p}-\ce{SiC} region after a sudden rise/drop of \textit{V}\textsubscript{DS}. 
 (d) Extrapolated effective time constant (\textit{\texttau}\textsubscript{eff}) of (dis)charging the \textit{p}-\ce{SiC} region. 
 Fixed \textit{N}\textsubscript{A} = 0.6$\times$10\textsuperscript{17} cm\textsuperscript{-3}, \textit{t}\textsubscript{p} = 0.6 \textmu m, and \textit{L}\textsubscript{p} = 7 \textmu m.
 The contact resistance of the Ohmic contact to the \textit{p}-\ce{SiC} is calibrated to 10\textsuperscript{-4}~\textOmega$\cdot$cm\textsuperscript{-2}~\cite{Huang2020critical}.
 The hole mobility in \ce{SiC} is estimated to be 100~cm\textsuperscript{2}$\cdot$V\textsuperscript{-1}$\cdot$s\textsuperscript{-1} according to the \textit{N}\textsubscript{A}~\cite{Matsuura2004Dependence}. 
 \textit{V}\textsubscript{GS} = $-$20~V.
 }
 \label{fig:fig8}
\end{figure}

During practical switching process, $V_\mathrm{DS}$ switches between high and low values.
When the $V_\mathrm{DS}$ switches to high values, holes will leave the $p$-\ce{SiC} to establish a depletion region (Fig.~\ref{fig:fig8}~(a)).
When the $V_\mathrm{DS}$ switches to low values, holes will flow back to the $p$-\ce{SiC}.
The charging/discharging rate of the depletion region is limited by the contact resistance and sheet resistance of the $p$-\ce{SiC}, and should be fast enough for high frequency applications.
To estimate the (dis)charging time of the $p$-\ce{SiC}'s depletion region, the simulation in Fig.~\ref{fig:fig8}~(b) is conducted.
First, $V_\mathrm{DS}$ is suddenly (within 0.1~ns) increased from 0 to 800~V (charging), or decreased from 800 to 0~V (discharging). 
Then, the current through the $p$-\ce{SiC} is monitored (Fig.~\ref{fig:fig8}~(c)),
from which the effective time constant ($\tau_\mathrm{eff}$) for charging/discharging are both estimated to be $\sim$0.5~ns after a $\sim$0.5~ns of oscillation (Fig.~\ref{fig:fig8}~(d)). 
Therefore, the (dis)charging time (usually defined at 5$\tau_\mathrm{eff}$) for the $p$-\ce{SiC}'s depletion region is $\sim$2.5~ns.
The fast (dis)charging rate demonstrates the feasibility of such $p$-\ce{SiC} RESURF structures in high frequency (MHz level) switching applications.

\section{Conclusion} \label{sec:conclusion}

In this work, the electrical properties of heat-dissipating \ce{Ga2O3}-on-\ce{SiC} heterosubstrate have been utilized to construct RESURF structures in \ce{Ga2O3} FETs.
The proposed RESURF structures can evenly distribute the $E$-field to achieve higher BVs.
With careful design of the $N_\mathrm{A}$ and dimensions of $p$-SiC, as well as eliminating the interface charges, the BV can be improved from 253~V to 2055~V with the RESURF structures. 
The \ce{Al2O3} interlayer in the existing heterosubstrate devices has minimal influence on the RESURF structures. 
Additionally, RESURF \ce{Ga2O3}-on-\ce{SiC} FETs shows a fast (dis)charging rate for high-frequency switching applications. 
This study provides a demonstration of unlocking the full potential of heat-dissipating heterosubstrates by leveraging their electrical properties.





\end{document}